\documentclass[prl,aps,twocolumn,amsmath,amssymb,showpacs]{revtex4}
\usepackage{graphicx}
\begin{document}
\title{Hydrodynamic singularities and clustering in a freely cooling inelastic gas}
\author{Efi Efrati, Eli Livne and Baruch Meerson}
\affiliation{Racah Institute of Physics, Hebrew University of
Jerusalem, Jerusalem 91904, Israel}
\begin{abstract}
We employ hydrodynamic equations to follow the clustering
instability of a freely cooling dilute gas of inelastically
colliding spheres into a well-developed nonlinear regime. We
simplify the problem by dealing with a one-dimensional
coarse-grained flow. We observe that at a late stage of the
instability the shear stress becomes negligibly small, and the gas
flows solely by inertia. As a result the flow formally develops a
finite time singularity, as the velocity gradient and the gas
density diverge at some location. We argue that flow by inertia
represents a generic intermediate asymptotic of unstable free
cooling of dilute inelastic gases.

\end{abstract}
 \maketitle

A gas of inelastically colliding macroscopic particles is a simple
paradigm of granular matter
\cite{Haff,Campbell,Kadanoff1,Goldhirsch1}, and it appears in
numerous applications, from astrophysics and geophysics to
materials processing. One of the many fascinating phenomena in a
freely cooling gas of inelastic particles is \textit{clustering
instability}
\cite{Hopkins,Goldhirsch2,McNamara,Barrat,Ernst,Brey,Luding,Ben-Naim2,plasma}.
This instability attracted much attention from physicists to rapid
granular flow \cite{Kadanoff1,Jaeger}. The clusters form an
intricate cellular structure
\cite{Goldhirsch2,McNamara,Barrat,Ernst,Luding}. Though molecular
dynamics (MD) simulations provide a valuable insight into the
complicated dynamics of clustering, a better understanding
requires a continuum theory. In this Letter we consider a freely
cooling dilute gas of identical inelastic hard spheres. In this
case a continuum theory is derivable systematically from a kinetic
equation, leading to hydrodynamic equations with a heat loss term
caused by the inelasticity of particle collisions
\cite{Haff,Jenkins,Sela,Brey1}. Hydrodynamics is expected to be an
accurate leading order theory when 
the mean free path of the particles is much less than any length
scale, and the mean time between two consecutive collisions is
much less than any time scale, described hydrodynamically.
Linearizing the hydrodynamic equations around a homogeneous
cooling state (HCS), one finds two different linearly unstable
modes: the shear mode and the thermal, or clustering, mode
\cite{Goldhirsch2,McNamara,Barrat}. Growth of the shear mode
corresponds to production of vorticity, while the clustering mode
governs cluster formation. Nonlinear evolution of the clustering
instability is a hard problem. Firstly, one has to deal here with
nonlinear coupling of the shear and clustering modes. Secondly, as
the local density grows, the hydrodynamic description becomes less
accurate. It breaks down completely when the density approaches
the point of the disorder-order transition in the gas of hard
spheres.

In this Letter we follow the clustering instability into a
well-developed nonlinear stage by circumventing these two
difficulties. Firstly, we put the particles into a long
two-dimensional (2D) box, $L_x \gg L_y$, so that the shear modes
are strongly over-damped, and all coarse-grained quantities depend
only on $x$. Secondly, we consider the limit of a very small area
fraction of the particles. In this case, despite clustering, the
gas density remains small, compared with the freezing density, for
a very long time. Importantly, this limit does not preclude
arbitrarily high
density contrasts in the system. 
We solve the hydrodynamic equations numerically and observe that,
at a late stage of the dynamics, the shear stress becomes
negligibly small. As a result, the gas moves only by inertia, and
the flow formally exhibits a finite time singularity. This
singularity has a universal character if the initial mean velocity
profile is smooth. We argue that flow by inertia is a generic
intermediate asymptotic in more general multi-dimensional freely
cooling granular flows, and that the finite time singularities
form the skeleton of the later dynamics, when finite-density
effects in the clusters come into play.

Let each of $N$ hard disks have a diameter $\sigma$ and mass
$m=1$. Let the inelasticity of particle collisions be
$q=(1-r)/2>0$, where $r$ is the (constant) coefficient of normal
restitution. Hydrodynamics deals with three coarse-grained fields:
the number density $n(x,t)$, the mean velocity $v(x,t)$ and the
granular temperature $T(x,t)$. We employ scaled variables $n
\rightarrow n/n_0$, $T \rightarrow T/T_0$, $v \rightarrow
v/T_0^{1/2}$, $x\rightarrow x/L_x$ and $t\rightarrow t\,
T_0^{1/2}/L_x$, where $n_0=N/(L_x L_y)$ and $T_0$ are the average
number density and the initial temperature of the gas,
respectively. In the dilute limit, granular hydrodynamic equations
\cite{Haff,Jenkins,Sela,Brey1} read:
\begin{equation}
\label{govern1} \frac{d n}{dt} + n \frac{\partial v}{\partial x} =
0 \,,\,\,\,\,\, \mbox{(a)} \;\;\;\;\; n \frac{d v}{dt} -
\frac{\partial P}{\partial x} =0 \,, \,\,\,\,\, \mbox{(b)}
\end{equation}
\begin{equation}
n \frac{d T}{dt} = P \frac{\partial v}{\partial x} + K
\frac{\partial}{\partial x}\left(T^{1/2}\frac{\partial T
}{\partial x} \right) - \frac{8q}{K}\,n^{2}T^{3/2}\,,
\label{govern3}
\end{equation}
where $d/dt = \partial/\partial t + v \,\partial/\partial x$ is
the total time derivative, $K=(2/\sqrt{\pi}) (\sigma L_x
n_0)^{-1}$ is the Knudsen number which, up to a constant factor of
order unity, is the ratio of the mean free path of the particles
to $L_x$, and $P=-n\,T + (K/4)\,T^{1/2} (\partial v/\partial x)$
is the stress field. The validity of Eqs. (\ref{govern1}) and
(\ref{govern3}) requires $K \ll 1$ (scale separation), $n \sigma^2
\ll 1$ (dilute limit), and $q \ll 1$ (nearly elastic collisions).

The HCS is described by Haff's cooling law
$T(x,t)=(1+t/t_0)^{-2}$, where $t_0=K/4q$ \cite{Haff}. A linear
stability analysis of the HCS, analogous to that of Refs.
\cite{Goldhirsch2,McNamara,Barrat}, predicts clustering
instability of the HCS if $K k_x < 2 q^{1/2}$, where $k_x$ is the
(scaled) wave number of a small sinusoidal perturbation around the
HCS. The wave number is quantized by the periodic boundary
conditions: $k_x = 2 \pi k$, where $k=1,2, \dots$ is the mode
number. Therefore, the $k$-th mode is linearly unstable if $\pi k
K<q^{1/2}$ \cite{noshear}. The rest of the parameters fixed, the
instability occurs when $L_x$ is sufficiently large. The number of
the unstable modes in the system $k_{\max}$ can serve as a measure
of the instability magnitude. The growth/decay of small
perturbations is algebraic. The density perturbations grow. The
temperature and velocity perturbations decay, but the decays are
slow compared to Haff's cooling law. As a result, the flow tends
to become supersonic \cite{Barrat}.

We followed the clustering instability with $k_{max} \gg 1$ into a
strongly nonlinear regime by solving Eqs. (\ref{govern1}) and
(\ref{govern3}) numerically. We used a Lagrangian scheme \cite{RM}
with periodic boundary conditions. The Lagrangian description
allowed us to resolve steep velocity gradients and high density
peaks with good accuracy until close to singularities, see below.
The first series of hydrodynamic simulations dealt with generic
initial conditions of the form $n(x,t=0) = 1 + \delta n(x)$,
$T(x,t=0)=1+ \delta T(x)$ and $v(x,t=0) = \delta v(x)$, where each
of the small terms $\delta n(x)$, $\delta T (x)$ and $\delta v(x)$
is a sum of a few hundred Fourier modes with random small
amplitudes, of which a few dozen modes are linearly unstable.  In
all these simulations we observed strong clustering: development
of multiple high and narrow density peaks, accompanied by
steepening velocity gradients, as the gas temperature continues to
decay. The gas density in the peaks grows without limit, until the
time when our finite-difference scheme is unable to accurately
follow the density growth in the highest density peak. The
temporal growth of the density peaks, and of the velocity
gradients, accelerates rapidly, implying a finite-time
singularity. Figure 1 shows a typical snapshot of the system close
to the time of singularity.

\begin{figure}
\includegraphics[width=9.5cm,clip=]{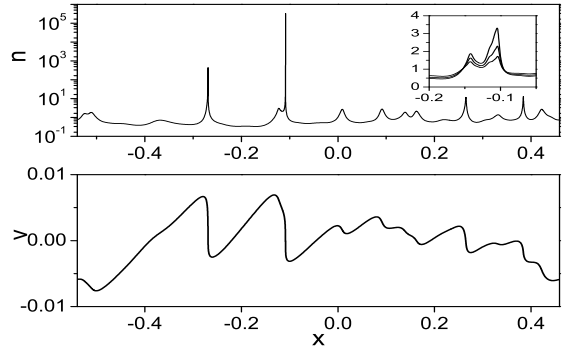}\vspace{-1cm}\caption{The
density and velocity profiles at scaled time $t=7.043$, shortly
before the major density peak develops singularity. The parameters
$K=4 \cdot 10^{-4}$ and $q=10^{-2}$ correspond to 79 linearly
unstable Fourier modes. $2000$ Lagrangian mesh points are used, so
the major density peak includes more than $50$ mesh points above
the density value of $n=10^2$. The inset shows an earlier density
history (at $t=2$, $3$ and $4$) of a region around the major
density peak.} \label{peaks}
\end{figure}

A convenient integral measure of the unstable cooling dynamics is
the total energy of the system:
\begin{equation}\label{total_energy}
E(t)=\int_{-1/2}^{1/2}\left(n\, T+\frac{1}{2}\, n \,v^2\right)
dx\,,
\end{equation}
where $n\,T$ is the thermal energy density, and $n\,v^2/2$ is the
macroscopic kinetic energy density. A plot of $E(t)$ is shown in
Fig. 2. As expected, $E(t)$ follows Haff's law at early times, but
deviates from it at later times. Figure 2 elucidates the role of
each of the two terms in Eq. (\ref{total_energy}). Both the
thermal energy, and the macroscopic kinetic energy initially decay
with time; the thermal energy decays faster. At later times the
kinetic energy approaches a constant. As a result, $E(t)$ is
dominated by the thermal energy at early times and by the kinetic
energy at later times. Remarkably, the thermal energy continues to
follow Haff's law until the time of singularity.

\begin{figure}[ht]
\includegraphics[width=5.5 cm,clip=]{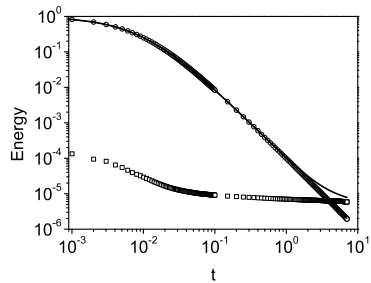}\caption{The total energy
of the system $E$ (thick solid line), the thermal energy
(circles), the macroscopic kinetic energy (squares) and Haff's
cooling law (thin solid line) versus time for the simulation shown
in Fig. 1.} \label{energy}
\end{figure}

The finite-time singularities of the velocity gradient and the
density strongly indicate that, at late times, the stress tensor
$P$ becomes negligibly small, and the gas flows by inertia only.
An additional evidence for the flow by inertia is the constancy of
the macroscopic kinetic energy at late times. The flow by inertia
is described by the equation
\begin{equation} \label{inertia}
\frac{\partial v}{\partial t} + v\,\frac{\partial v}{\partial x} =
0
\end{equation}
and Eq. (\ref{govern1})a. This problem is soluble analytically
\cite{Whitham}:
\begin{equation}\label{solution}
v(x,t) = v_0(\xi)\,,\,\,\,\,(a) \;\;\;\;\;\;
n(x,t)=\frac{n_0(\xi)}{1+t\,v_0^{\prime}(\xi)}\,. \,\,\,\,\,(b)
\end{equation}
where $v_0^{\prime}(\xi)= d v_0(\xi)/d \xi$, while $v_0(\xi)$ and
$n_0(\xi)$ are the velocity and density of the gas, respectively,
at some ``initial" moment of time (which should be late enough so
that the flow by inertia has already set in). The relation between
Eulerian coordinate $x$ and Lagrangian coordinate $\xi$ is the
following: $x=\xi+ v_0(\xi)\,t$.
The finite-time singularities of both the velocity gradient
\begin{equation}\label{v_gradient}
\frac{\partial v(x,t)}{\partial x} =
\frac{v_0^{\prime}(\xi)}{1+t\,v_0^{\prime}(\xi)}\,,
\end{equation}
and the density, Eq. (\ref{solution})b,  occur when the
denominator in Eq. (\ref{v_gradient}) becomes zero for the first
time. We compared these predictions with a numerical solution of
the full hydrodynamic equations (\ref{govern1}) and
(\ref{govern3}), for the same parameters $K=4 \cdot 10^{-4}$ and
$q=10^{-2}$, but with simpler initial conditions:
$n(x,t=0)=T(x,t=0)=1$, and a single Fourier mode for the velocity:
\begin{equation}\label{simple}
v(x,t=0)= a \sin (2\pi x)\,,\;\;\;a=-0.05\,.
\end{equation}
In this case only one singularity develops (at $x=0$). Figure 3
shows the gas velocity $v$ versus $\xi = x-t\,v(x,t)$ at different
times. The different curves collapse into a single curve with an
accuracy better than 1.5\%.
Additional tests deal with the behavior of the flow in the close
vicinity of $x=0$, as the singularity time is approached. For this
smooth symmetric flow we can write $v_0(\xi) = - \xi/\tau + C
\xi^3+ {\cal O}(\xi^5) $, where $t=\tau$ is the time of
singularity in the flow-by-inertia model, and $C>0$ is a constant.
In the Eulerian coordinates this yields a solution in an implicit
form. In the leading order
\begin{equation}\label{implicit}
x = - t^{\prime} \,v (x,t^{\prime}) -C \tau^4 v^3(x,t^{\prime})
\,,
\end{equation}
where $t^{\prime}=\tau-t$ is the time to the singularity. Not too
close to the singularity point $x=0$ one obtains $v \sim
(-x)^{1/3}$. As the velocity profile (\ref{implicit}) is
self-similar: $v(x,t^{\prime})= (t^{\prime})^{1/2} \,
V\left[x/(t^{\prime})^{3/2}\right]$, the velocity gradient is
$\partial v/\partial x = (t^{\prime})^{-1} dV/dw$, where
$w=x/(t^{\prime})^{3/2}$. The shape function $V(w)$ is determined
by the equation $C\tau^4 V^3+V+w=0$. What is the density behavior
close to the singularity? Very close to $x=0$ the density grows
indefinitely: $n_0(0) (1-t/\tau)^{-1}$; outside of that region
(but still close enough to $x=0$) $n(x,t)$ approaches a universal
profile $n \sim |x|^{-2/3}$ \cite{Zeldovich1}. We verified these
properties numerically, see examples in Fig. 4. Importantly, for a
strong instability, $k_{max} \gg 1$, the system ``freezes up", and
the motion by inertia sets in very rapidly. Indeed, the scaled
velocity profile in Fig. 3 is very close to the \textit{initial}
profile (\ref{simple}). Building on this simplification, we can
expand Eq. (\ref{simple}) in the vicinity of $x=0$ and predict the
time of singularity: $\tau = (2 \pi |a|)^{-1} \simeq 3.18$ which
agrees within 2\% with the simulation result, see Fig. 4a. In
addition, the linear time dependence of the quantities, shown in
the inset of Fig. 4a, sets in already at early times.

\begin{figure}
\includegraphics[width=7.4 cm,clip=]{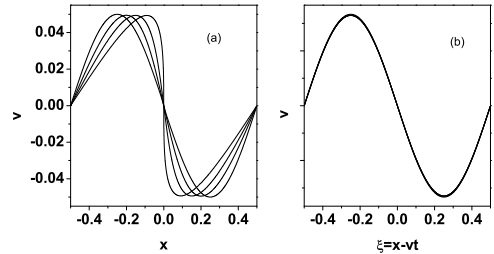}\vspace{-0.5cm}\caption{The numerically computed velocity
is shown versus $x$ (a) and versus $\xi = x-v\,t$ (b) at times
$1$, $2$ and $3.225$ (the profiles in figure a steepen as the time
progresses). Also shown in figure b is the initial profile
(\ref{simple}). All the curves in figure b coincide within 1.5\%.
The simulation parameters are $K=4 \cdot 10^{-4}$ and
$q=10^{-2}$.}
\end{figure}

Therefore, a strongly nonlinear regime of the
\textit{quasi}-one-dimensional clustering instability in a dilute
granular flow is describable by a simple flow by inertia, until
the moment of singularity \cite{MD}. In a related work Ben-Naim
\textit{et al.} \cite{Ben-Naim1} investigated the dynamics of
\textit{point-like} particles, inelastically colliding on a
\textit{line}. The strictly one-dimensional setting of Ref.
\cite{Ben-Naim1} makes a hydrodynamic description problematic.
Still, Ben-Naim \textit{et al.} observed that the Burgers equation
with vanishing viscosity is a proper continuum model for their
system. It remains to be seen whether the Burgers equation or some
other saturation mechanism applies to our
\textit{quasi}-one-dimensional model at a later stage of the
dynamics, when finite-density effects come into play. We stress
that the (hydrodynamic) density singularities are entirely
different from \textit{inelastic collapse} \cite{McNamara2}
(divergence of the particle collision rate at some locations)
which is a discrete-particle effect. We also note in passing that
\textit{statistical} properties of the flow by inertia  (for
example, the dynamics of the structure function) are well
understood \cite{Gurbatov}.
\begin{figure}
\includegraphics[width=9.3 cm,clip=]{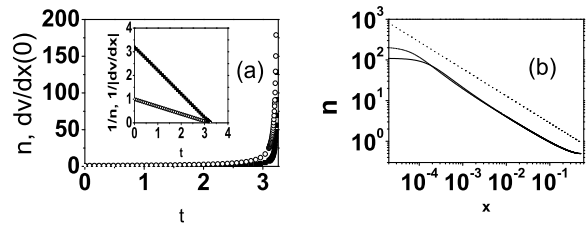}\vspace{-1cm}\caption{
Numerically computed values of $|\partial v/\partial x|$ (filled
squares) and $n$ (empty circles) at x=0 versus time (a). The inset
shows the respective inverse values. Figure b depicts the spatial
profiles of $n$ at time moments $3.209$ and $3.225$. The straight
line is a $x^{-2/3}$ dependence; it is given for reference. The
parameters are the same as in Fig. 3.}
\end{figure}

What can be said about a fully multi-dimensional strongly unstable
cooling flow, when unstable shear and clustering modes are
coupled? A natural assumption, motivated already by the linear
theory of the clustering/shearing instabilities of the HCS
\cite{Goldhirsch2,McNamara,Barrat}, is that the stress tensor
``freezes up", and flow by inertia sets in here as well. A
possible counter-argument involves viscous heating of the system
by the unstable shear modes. The heating effect is absent in the
linear regime of the instability (as the viscous heating is of the
second order with respect to the perturbation amplitude), but it
comes into play in the nonlinear regime. The present state of
theory makes it difficult to prove that the viscous heating cannot
arrest, in some locations, the freezing of the stress tensor.
However, MD simulations in 2D strongly indicate that the freezing
continues unarrested.  For example, Nie \textit{et al.}
\cite{Ben-Naim2} observe that, at late times, ``the thermal energy
becomes much smaller than the (macroscopic) kinetic energy". Based
on this evidence we argue that, in the dilute regime, this
strongly supersonic flow should be describable by
\textit{multi-dimensional} flow-by-inertia equations
\begin{equation}\label{multi-free}
\partial \mathbf{v}/\partial t + (\mathbf{v} \cdot
\mathbf{\nabla}) \mathbf{v}=\mathbf{0}\,, \,\,\,\,\,\,\,\,\,\,
\partial n /\partial t+ \mathbf{\nabla} \cdot (n \mathbf{v}) =0.
\end{equation}
This flow also exhibits finite-time singularities
\cite{Zeldovich1,Arnold-Bruce}, and the singularities form
cellular structures, most of the material being concentrated along
the cell boundaries \cite{Zeldovich1}. This picture resembles the
density distribution of granular clusters observed in MD
simulations of freely cooling gases of inelastic hard spheres in
2D \cite{Goldhirsch2,McNamara,Barrat,Ernst,Luding}. Interestingly,
the multi-dimensional singularities of Eqs. (\ref{multi-free})
were studied previously in an entirely different context: in the
so called Zeldovich approximation of theory of formation of
structure in an expanding universe \cite{Zeldovich1}.

We stress that there are important differences between the
multi-dimensional clustering instability and the Zeldovich model.
In the process of clustering instability of inelastic gases a
considerable vorticity is generated, while in Zeldovich model the
flow is assumed to be potential \cite{Zeldovich1}. Still, it was
found, in a rare treatment of a more general (non-potential)
velocity field, that ``high-density regions should be
high-vorticity regions" \cite{Barrow}. This finding appears to
agree with MD simulations of freely cooling granular gases in 2D
\cite{McNamara}.

In summary, by following the unstable cooling dynamics of a
dilute inelastic gas we identified an important new intermediate
asymptotic regime: a nonlinear flow by inertia. We argue that
high-density regions in the gas, which are precursors of densely
packed granular clusters, are caused by the flow by inertia,
rather than directly by the pressure gradient. Our results
indicate that the role of the clustering and shearing
instabilities of the free cooling is ``merely" to produce a
long-lived spatially non-uniform supersonic velocity field needed
for the development of the high-density regions by the flow by
inertia. Therefore, a due account of the flow-by-inertia regime
will be important in the future theory of ``life after
singularity", where the singularities are smoothed by
finite-density effects in the clusters, and a coarsening process
develops \cite{Luding}. No first-principles coarse-grained
description of that final stage is yet available.

We thank A. Puglisi, P. V. Sasorov and S. F.  Shandarin for useful
discussions. This research was supported by the Israel Science
Foundation.


\begin{references}
\bibitem{noshear} To suppress shear modes and clustering modes in the $y$-direction,
we impose reflecting boundary conditions at $y=0$ and $y=L_y$ and
demand $q^{1/2} \ll (\pi/4) K (L_x/L_y)$.
\bibitem{MD} The quasi-one-dimensional regime can be probed
in MD simulations in 2D. The parameters must be chosen so that the
gas is dilute, the clustering instability in the $x$-direction is
strong, and the shear/clustering instabilities in the
$y$-direction are strongly suppressed
\cite{noshear}. 
\end{references}
\end{document}